\documentclass[11pt,a4paper]{article}

\usepackage{amsmath}
\usepackage{vmargin}
\usepackage{amsfonts}
\usepackage[parfill]{parskip}
\usepackage[labelfont=bf]{caption}
\usepackage{graphicx}
\usepackage[round,comma]{natbib}
\usepackage{bbm}

\newcommand{\be}{\begin{equation}}
\newcommand{\ee}{\end{equation}}
\newcommand{\ba}{\begin{equation} \begin{aligned}}
\newcommand{\ea}{\end{aligned} \end{equation}}
\newcommand{\ddt}[1]{\frac{\mathrm{d}#1}{\mathrm{d}t}}

\newcommand{\dint}[1]{\mathrm{d}#1}
\newcommand{\myvec}[1]{ \mathbf{#1} }
\newcommand{\mymat}[1]{ \mathbf{#1} }
\newcommand{\mygmat}[1]{ \boldsymbol{#1} }
\newcommand{\mygvec}[1]{ \boldsymbol{#1} }
\newcommand{\unit}{\mathbbm{1}}

\newcommand{\TR}[1]{#1^{\!\top}}
\newcommand{\INV}[1]{#1^{-1}}

\title{\bf Hessian corrections to the Metropolis Adjusted Langevin Algorithm}
\author{Thomas House\\
\textit{School of Mathematics, University of Manchester, Manchester, M13 9PL, UK.}}
\date{}

\begin{document}

\maketitle

\begin{abstract}
	\noindent{}A natural method for the introduction of second-order derivatives
	of the log likelihood into MCMC algorithms is introduced, based on Taylor
	expansion of the Langevin equation followed by exact solution of the
	truncated system.
\end{abstract}

\section{Introduction}

Markov chain Monte Carlo (MCMC) is a highly influential computationally
intensive method for performing Bayesian inference, with a large variety of
applications \citep{Brooks:2011}.  While earlier MCMC algorithms made use of
random walks in parameter space \citep{Gilks:1995}, as highlighted in a recent
review by \citet{Green:2015}, the use of derivatives can lead to improved
algorithms.

One derivative-based approach is the Metropolis-adjusted Langevin Algorithm,
MALA \citep{Roberts:1996a}, which requires first derivatives of the log
likelihood to be available.  More recently, second derivatives have been
included via the use of geometric approaches \citep{Girolami:2011}, the
MALA-like versions of which are often highly efficient in applications
\citep{Calderhead:2011,Kramer:2014}. Other approaches include more general
position-dependent MALA (PMALA, analysed by \citet{Xifara:2014}) although the
tuning of these in the absence of an appropriate metric for geometric
approaches remains a problem.

This letter introduces a different route to inclusion of second-order
derivatives through truncated Taylor expansion of the log-likelihood, after
which the Langevin equation can be solved exactly without further
approximation. This algorithm is called HMALA (for Hessian-corrected MALA) and
leads to a four-fold improvement on the effective sample size compared to
random walk approaches for a simple example, as well as being able to deal with
non-convex distributions.

\section{A Hessian MALA algorithm}

\subsection{Local solution of the Langevin equation}

From \citet{Roberts:1996b}, we know that the following Langevin SDE has stationary
distribution $\pi$ (subject to technical conditions):
\be
\dint{\mygvec{\theta}} = \frac{1}{2}\nabla \mathrm{ln}(\pi(\mygvec{\theta}))
\dint{t} + \dint{\myvec{W}} \text{ .}
\ee
Now suppose that we approximate $l = \mathrm{ln}(\pi)$ in the neighbourhood of some value
$\mygvec{\theta}^n$ through Taylor expansion
\be
l(\mygvec{\theta}^n + \myvec{x}) \approx l(\mygvec{\theta}^n) +
\TR{\myvec{v}}\myvec{x} + \frac{1}{2} \TR{\myvec{x}}\mymat{H} \myvec{x} \text{ ,}
\ee
where
\be
v_i := \left. \frac{\partial l}{\partial \theta^i} \right|_{\mygvec{\theta}^n} \text{ ,} \quad
H_{ij} := \left. \frac{\partial^2 l}{\partial \theta^i\partial \theta^j}
\right|_{\mygvec{\theta}^n} \text{ ,} \quad
\myvec{v} := (v_i) \text{ ,} \quad
\mymat{H} := (H_{ij}) \text{ .}
\ee
Then we can approximate the Langevin SDE in the region of $\mygvec{\theta}^n$ through
the linear SDE
\be
\dint{\myvec{x}} = \frac{1}{2} \left(
\mymat{H}\myvec{x} + \myvec{v}\right)
\dint{t} + \dint{\myvec{W}} \text{ .}
\ee
From the results of \citet{Archambeau:2007}, this linear SDE has Gaussian
solution with mean $\myvec{m}$ and covariance matrix $\mymat{S}$ obeying
\be
\ddt{\myvec{m}} = \frac{1}{2}\left(\mymat{H}\myvec{m} + \myvec{v}
\right) \text{ ,} \qquad
\ddt{\mymat{S}} = \mymat{H}\mymat{S} + \unit \text{ .} \label{mSODE}
\ee
Solving these ODEs over the interval $[0, \delta]$ with initial conditions
$\myvec{m}(0) = \mathbf{0}$, $\mymat{S}(0) = \mathbf{0}$, gives the solution
\be
\myvec{m} = \left({\rm e}^{\frac{1}{2}\mymat{H}\delta} - \unit \right)
\INV{\mymat{H}} \myvec{v} \text{ ,} \qquad
\mymat{S} = \left({\rm e}^{\mymat{H}\delta} - \unit \right)
\INV{\mymat{H}}\text{ .} \label{sol}
\ee
This solution is best understood in terms of the power-series definition of the
matrix exponential
\be
{\rm e}^{\mymat{M}} = \sum_{a=0}^{\infty} \frac{\mymat{M}^a}{a!} 
\quad \Rightarrow \quad
\phi_1 (\mymat{M}) = \left({\rm e}^{\mymat{M}} - \unit \right)
\INV{\mymat{M}} = \sum_{a=0}^{\infty} \frac{\mymat{M}^a}{(a+1)!} \text{ .}
\label{matdef}
\ee
Substituting~\eqref{matdef} into~\eqref{sol} gives a series that is clearly
the solution to~\eqref{mSODE}, subject to the initial conditions.

In terms of numerical computation of~\eqref{sol}, various options are
available.  These include: (i) direct computation of matrix exponentials and
inverses using e.g.\ \texttt{expm()} and \texttt{inv()} in MATLAB; (ii)
solving~\eqref{mSODE} using standard methods for ODEs such as Runge-Kutta;
(iii) use of numerical methods for matrix functions to calculate $\phi_1$,
which is quite well studied with a recent example being the methods of
\citet{Niesen:2012}. For the examples considered below, (i) performed well,
however it is likely that either (ii) or (iii) would be preferable for higher
dimensional problems.

\subsection{Metropolis-Hastings scheme}

The proposal density for HMALA is then
\be
q(\mygvec{\theta}^*| \mygvec{\theta}^n) = \mathcal{N}(\mygvec{\theta}^*
| \myvec{m} + \mygvec{\theta}^n; \mymat{S}) \text{ ,} \label{Hprop}
\ee
leading to acceptance probability
\be
\alpha = 1 \wedge
\frac{\pi(\mygvec{\theta}^*)q(\mygvec{\theta}^n| \mygvec{\theta}^*)}%
{\pi(\mygvec{\theta}^n)q(\mygvec{\theta}^*| \mygvec{\theta}^n)} \text{ .}
\ee
Standard MALA is recovered from HMALA by using~\eqref{Hprop} at first order in
$\delta$:
\be
\myvec{m} = \frac12 \myvec{v} \delta + o(\delta) \text{ ,} \qquad
\mymat{S} = \delta \unit + o(\delta)\text{ .}
\ee
For the random-walk (RW) algorithm, we ignore gradient information entirely
and use proposal density
\be
\tilde{q}(\mygvec{\theta}^*| \mygvec{\theta}^n) = \mathcal{N}(\mygvec{\theta}^*
| \mygvec{\theta}^n; \delta \unit) \text{ .}
\ee
It is worth noting in general that the solution~\eqref{sol} has some
similarities with the matrix cosh form suggested by \citet{Betancourt:2013} for
a metric in geometric approaches, $({\rm e}^{\mymat{\alpha \mymat{H}}} + {\rm
e}^{\mymat{-\alpha \mymat{H}}}) \mymat{H} ({\rm e}^{\mymat{\alpha \mymat{H}}} +
{\rm e}^{\mymat{-\alpha \mymat{H}}})^{-1}$. The important differences are,
however, that: (i) HMALA is not mathematically equivalent to any existing
geometric approach; (ii) HMALA can be used when the Hessian cannot be
integrated over all data meaning the Fisher-Rao metric is not available; (iii)
HMALA does not require tuning an additional parameter $\alpha$ as in the matrix
cosh approach.

\section{Examples}

\subsection{Negative binomial counts}

\begin{figure}[H]
\centering
\includegraphics[width=0.98\textwidth]{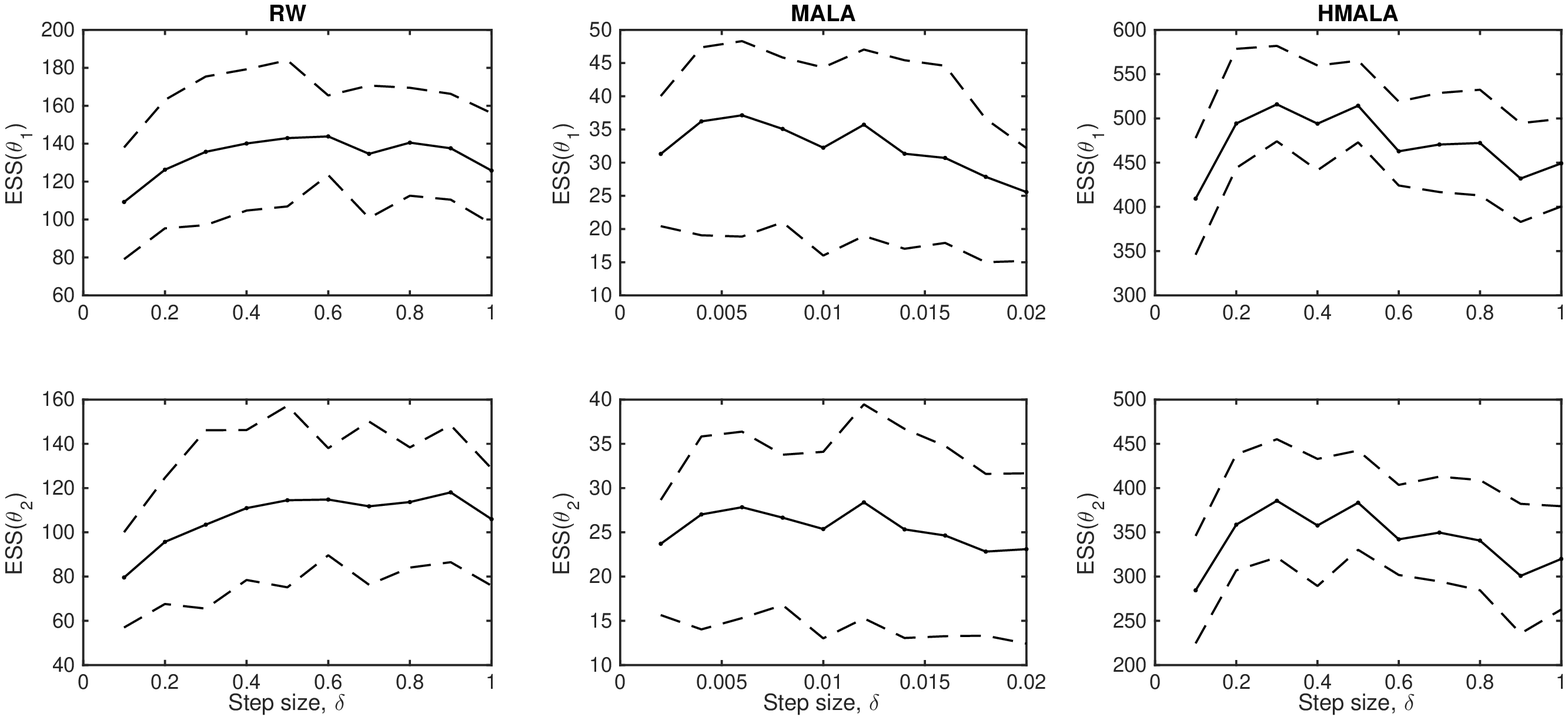}
\caption{Mean and 50\% CI for the effective sample size versus step size
	$\delta$ for the three algoritms. In each case 100 chains of length $10^4$
	were run.
\label{fig:ess}}
\end{figure}

Consider sampling from a density proportional to the likelihood function for a
model of $n$ negative binomial distributed integers, represented as a vector
$\myvec{k} = (k_a)$, leading to
\be
l(\myvec{k} | p, r) = \sum_{a=1}^{n} \left(
	\mathrm{ln}(\Gamma(k_a +r)) - \mathrm{ln}(k_a!) + k_a \mathrm{ln}(p)
	- \mathrm{ln}(\Gamma(r)) + r \mathrm{ln}(1-p) 
\right)\text{ ,}
\ee
\be
\partial_p l = \sum_{a=1}^{n} \left(
	\frac{k_a}{p} - \frac{r}{1-p}
\right)\text{ ,} \qquad
\partial_r l = \sum_{a=1}^{n} \left( \psi_1(k_a + r)
	- \psi_1(r) + \mathrm{ln}(1-p) 
\right)\text{ ,}
\ee
\be
\partial^2_p l = -\sum_{a=1}^{n} \left(
	\frac{k_a}{p^2} + \frac{r}{(1-p)^2}
\right)\text{ ,} \quad
\partial_p \partial_r l = \frac{-n}{1-p} \text{ ,}\quad
\partial^2_r l = \sum_{a=1}^{n} \left( \psi_2(k_a + r)
	- \psi_2(r) \right)\text{ .}
\ee
In this example, the Hessian is easily computed, but its expected value over
all data (needed to calculate the Fisher-Rao metric) involves infinite sums
that do not have known closed forms. To produce a likelihood function, 100
integers were simulated with `true' parameters $r = \theta_1 = 1.5$ and $p =
\theta_2 = 0.4$. Each of the algorithms RW, MALA and HMALA defined above was
run on this likelihood function. Results of calculating the effective sample
size as defined by Neal in the discussion of~\citet{Kass:1998} are shown in
Figure~\ref{fig:ess}.

\begin{figure}[H]
\centering
\includegraphics[width=0.98\textwidth]{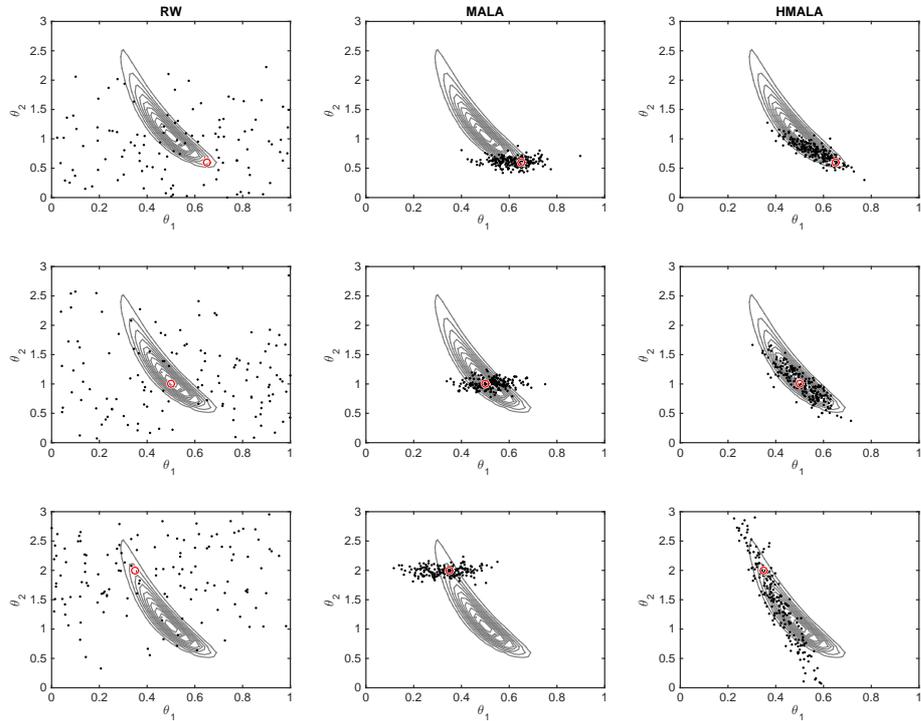}
\caption{200 proposals generated using different methods for the negative binomial count model,
for step sizes (RW) $\delta = 0.6$, (MALA) $\delta = 0.006$, (HMALA) $\delta =
0.5$, at three different parameter values.  \label{fig:props}}
\end{figure}

Figure~\ref{fig:props} shows how the different algorithms behave at the optimal
value of ESS. While RW is more efficient than MALA for this system, this is
primarily because in two dimensions ambitious proposals can be efficient, which
would not hold for more complex systems. MALA offers conservative local
proposals into relatively high-density regions, but HMALA is able to use
higher-order derivative information to make ambitious proposals into
high-density regions while achieving the largest ESS by a factor of about four.

\subsection{A Gaussian mixture}

\begin{figure}[H]
\centering
\includegraphics[width=0.98\textwidth]{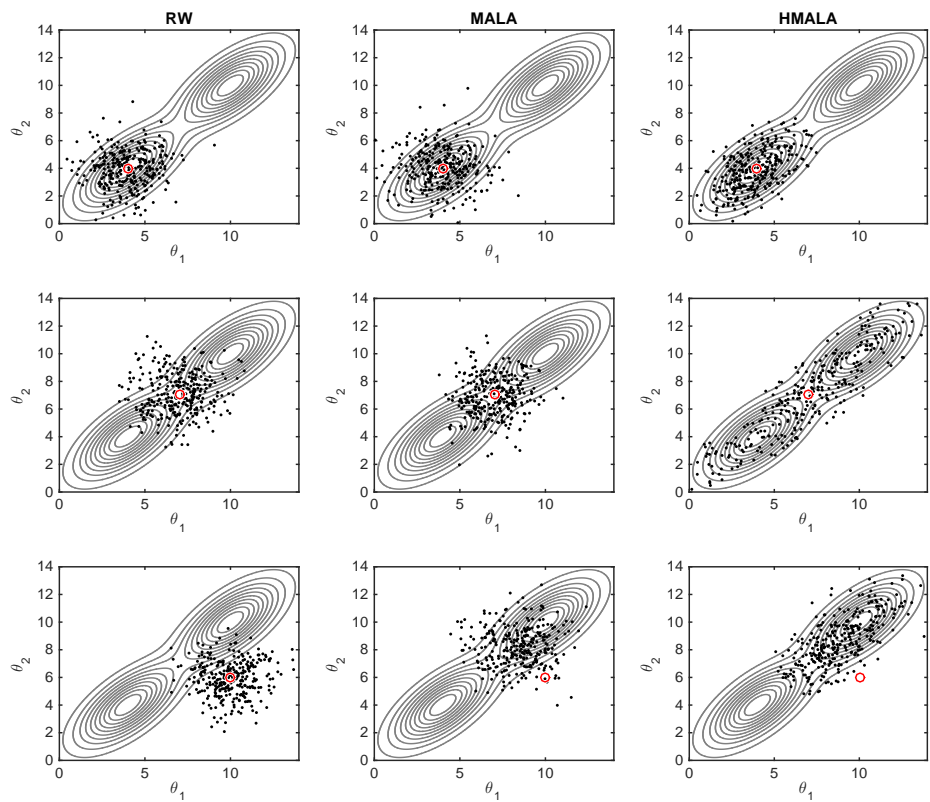}
\caption{300 proposals generated using different methods for the Gaussian mixture model,
for step sizes (RW) $\delta = 2$, (MALA) $\delta = 2$, (HMALA) $\delta =
6$, at three different parameter values.  \label{fig:conv}}
\end{figure}

Next, consider the following bimodal Gaussian mixture density:
\be
\pi(\mygvec{\theta}) = \frac12 \left(\mathcal{N}(\mygvec{\theta}
	| \mygvec{\mu}_1; \mygmat{\Sigma}) +\mathcal{N}(\mygvec{\theta}
	| \mygvec{\mu}_2; \mygmat{\Sigma}) \right) \text{;}
\quad
\mygvec{\mu}_1 = \begin{pmatrix}4\\4\end{pmatrix} \text{;}\quad
\mygvec{\mu}_2 = \begin{pmatrix}4\\4\end{pmatrix} \text{;}\quad
\mygmat{\Sigma} = \begin{pmatrix}3 & 2\\2 & 3\end{pmatrix} \text{ .}
\ee
This example exhibits bimodality, as well as a saddle point in a region of high
posterior density. As can be seen from Figure~\ref{fig:conv}, this does not
affect the ability of HMALA to propose efficient moves at the saddle point, the
modes, or in regions of low posterior density.

\section*{Acknowledgements}

Work supported by the UK Engineering and Physical Sciences Research Council. I
would like to thank Simon Cotter, Ashley Ford and Theo Kypraios for helpful
comments on this approach.

\end{document}